\documentclass[aps,superscriptaddress,floatfix,reprint]{revtex4-2}

\usepackage{graphicx,color}
\usepackage{amssymb,amsmath}
\usepackage[normalem]{ulem}

\begin{document}

\title{Internal energy exchanges and chaotic dynamics in an intrinsically coupled system}

\author{M. C. de Sousa}
\email[]{meirielenso@gmail.com}
\affiliation{Instituto de Fisica, Universidade de Sao Paulo, Sao Paulo, Brazil}
\affiliation{LSI, CEA/DRF/IRAMIS, CNRS, Ecole Polytechnique, Institut Polytechnique de Paris, F-91128 Palaiseau, France}
\author{A. B. Schelin}
\email[]{aschelin@gmail.com}
\affiliation{Instituto de Fisica, Universidade de Brasilia, Brasilia, Brazil}
\author{F. A. Marcus}
\email[]{albertus.marcus@gmail.com}
\affiliation{Instituto de Fisica, Universidade de Sao Paulo, Sao Paulo, Brazil}
\author{R. L. Viana}
\email[]{viana@fisica.ufpr.br}
\affiliation{Departamento de Fisica, Universidade Federal do Parana, Curitiba, Brazil}
\affiliation{Instituto de Fisica, Universidade de Sao Paulo, Sao Paulo, Brazil}
\author{I. L. Caldas}
\email[]{ibere@if.usp.br}
\affiliation{Instituto de Fisica, Universidade de Sao Paulo, Sao Paulo, Brazil}

\begin{abstract}
Intrinsically coupled nonlinear systems present different oscillating components that exchange energy among themselves. A paradigmatic example is the spring pendulum, which displays spring, pendulum, and coupled oscillations. We analyze the energy exchanges among the oscillations, and obtain that it is enhanced for chaotic orbits. Moreover, the highest rates of energy exchange for the coupling occur along the homoclinic tangle of the primary hyperbolic point embedded in a chaotic sea. The results show a clear relation between internal energy exchanges and the dynamics of a coupled system.
\end{abstract}

\maketitle

A challenging problem of intrinsically coupled nonlinear Hamiltonian systems is the internal energy exchange among their components \cite{Ford1992FermiPastaUlam}. One example is the numerical investigation of energy exchanges among normal modes in the Fermi-Pasta-Ulam-Tsingou (FPUT) model of a nonlinear coupled oscillators chain, which presents a surprising energy concentration in some modes \cite{Dauxois2005FermiPastaUlam, tsingou}. A reason for the observed results is the coexistence of regular and chaotic regions in the phase space of the coupled oscillators chain.

One of the difficulties related to the analysis of FPUT (and related) models is their large phase space dimension. In order to investigate with more depth the internal energy exchanges due to coupling between modes, we consider a paradigmatic low-dimensional system: the spring pendulum with two degrees of freedom \cite{Vitt1933OscillationsSystems}. This system exhibits an intrinsic nonlinear coupling, and we present numerical evidence that the rate of energy exchanges associated with the coupling is directly related to the regular or chaotic dynamics of the trajectories.

The spring pendulum, also known as elastic or extensible pendulum, is composed of a spring connected to a pivot on one end and to a suspended mass on the other end. The suspended mass can oscillate both harmonically, due to the spring elasticity, and pendularly, performing librations and rotations around the pivot due to the gravitational force. The combination of these two oscillation modes yields a complex and rich dynamics, with chaotic and resonant effects such as an order-chaos-order transition \cite{Nunez-Yepez1990OnsetPendulum, Cuerno_AmJP92, Gonzalez_EJP94, Weele_PhysA1996} and a parametric resonance \cite{Vitt1933OscillationsSystems, Kane_JAM1968, Tselman_JAMM1970, Rusbridge_AmJP1980, Breitenberger_JMP1981, Lai_AmJP1984} according to the total energy and system parameters. 

Besides having a complex dynamics, the spring pendulum is also investigated due to its similarities to other physical systems of great interest. Fermi already pointed out that the spring pendulum dynamics can be a mechanical analogy to the interaction of an atom with radiation \cite{Fermi1931UberSteinsalzes}. Fermi also brought to the attention that the parametric resonance in the vibrations of the two degrees of freedom $\text{CO}_2$ molecule is similar to the spring pendulum parametric resonance \cite{Fermi1931UberSteinsalzes, Amat1965OnDioxide, Jacob_JPB1978}. Among other examples of systems described by equations analogous to those of the spring pendulum, we mention the orbits of celestial bodies \cite{Contopoulos1963ResonanceI, Hori_ASJ1966, Broucke1973PeriodicSystem, Hitzl_CM1975}, coupled waves in plasmas \cite{Sagdeev1969NonlinearTheory, Horton2012TurbulentPlasmas}, interaction between light waves in a nonlinear dielectric \cite{Armstrong_PR1962}, and several mechanical devices \cite{Holmes2006TheChallenges, Anh2007VibrationAbsorber, Wang2011, Castillo-Rivera2017}.

In a previous work, we introduced three analytical expressions for the energy associated with the spring, pendulum and coupled oscillations of a spring pendulum \cite{C.deSousa2018EnergyDistrib}. We applied these analytical expressions and verified that they accurately describe the energy components for periodic, quasi-periodic and chaotic trajectories. We then numerically computed phase space and time averages to determine how the energy is distributed among the three components as a function of the total energy and a control parameter representing the ratio of the pendulum and spring frequencies.

In the present paper, we investigate the energy exchange rate (i.e. the power) for the previously identified energy components. We examine the phase space configuration for different values of total energy and control parameter, and we find that regular regions exhibit well defined power components, whereas chaotic regions display non-uniform values for the power components. Furthermore, the power associated with the coupling is generally higher for chaotic orbits than for regular ones. We also analyze the time dependence of the coupling power and verify that its maximum values in the Poincar\'e section follows the homoclinic tangle \cite{OzoriodeAlmeida1988HamiltonianSystems, Lichtenberg1992, Meiss1992}, which is composed of homoclinic intersections of the stable and unstable manifolds of the primary hyperbolic point immersed in a chaotic sea.

Such results show that the regular or chaotic dynamics observed for this nonlinear coupled system is determined by the rate of internal energy exchanges among its components, and that the coupling is closely related to the manifolds of hyperbolic points. In this sense, the approach we introduce may be a new and practical way to distinguish chaotic and regular orbits, as well as to access the manifolds of hyperbolic points with good precision.

\begin{figure}[!t]
    \centering
    \includegraphics[width=0.4\linewidth]{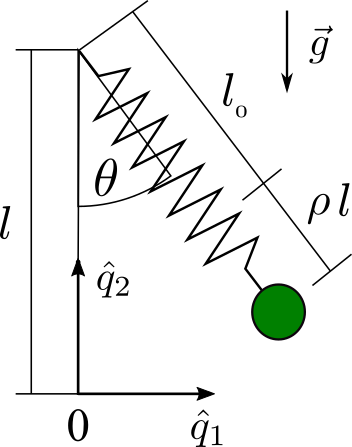}
    \caption{Spring pendulum diagram.}
    \label{fig:1}
\end{figure}

The vertical spring pendulum presents a mass $m$ connected to the free end of a spring with negligible mass, stiffness constant $k$ and relaxed length $l_0$ in the absence of forces, as shown in Figure \ref{fig:1}. The other spring end is fixed at the origin of the Cartesian coordinate system $(x,y) = (0,0)$. The spring pendulum moves in the vertical $xy$-plane and its stable equilibrium position is located at $x = 0$, $y = -l$, with $l = l_0 + mg/k$ and $g$ the acceleration of gravity.

We write the Hamiltonian of this system in the dimensionless Cartesian coordinates $q_1 = x/l$ and $q_2 = (y + l)/l$, which are centered in the stable equilibrium position, and their associated momenta $p_1 = {\dot q}_1$ and $p_2 = {\dot q}_2$:
\begin{eqnarray} \label{eq:Hq1q2}
	E_T = H & = & \frac{p_1^2 + p_2^2}{2} + f(q_2 - 1)     \nonumber  \\*
	        & + & \frac{1}{2}(\sqrt{q_1^2 + (q_2-1)^2} + f - 1)^2 ,
\end{eqnarray}
where $E_T$ is the dimensionless total energy, and $f = mg/(kl)$ is the square of the ratio between the pendulum and spring frequencies in the linear approximation, thus accounting for the physical characteristics of the spring pendulum.

\begin{figure}[!t]
    \centering
    \includegraphics[width=0.8\linewidth]{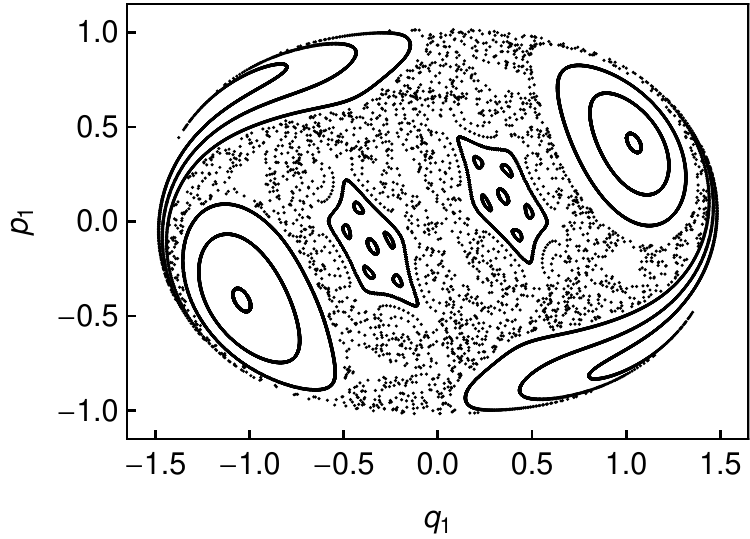}
    \caption{Poincar\'e section $p_1 \times q_1$, with $q_2 = 0$ and $p_2 > 0$, for $E_T = 0.3$ and $f=0.25$.}
    \label{fig:ps03025}
\end{figure}

As an example, we show in Figure \ref{fig:ps03025} the Poincar\'e section of the spring pendulum for $E_T=0.3$ and $f=0.25$, which displays a mixed dynamics typical of Hamiltonian systems, where islands of regular motion are embedded in the chaotic area. In fact, by increasing the total energy and/or control parameter, the Poincar\'e sections of the spring pendulum reveal a peculiar sequence of order-chaos-order \cite{Nunez-Yepez1990OnsetPendulum, Cuerno_AmJP92, Gonzalez_EJP94, Weele_PhysA1996}.

Following Ref. \cite{C.deSousa2018EnergyDistrib}, we work with the dimensionless polar coordinates
\begin{equation}
    \rho=f-1+\sqrt{q_1^2+(q_2-1)^2}
\end{equation}
and
\begin{equation}
    \tan \theta = \frac{q_1}{1-q_2},
\end{equation}
where $\rho$ represents the spring extension or compression from its relaxed length, and $\theta$ is the pendulum displacement from its equilibrium position.

The Hamiltonian in dimensionless polar coordinates is then given by \cite{C.deSousa2018EnergyDistrib}:
\begin{eqnarray} \label{eq:H_rho_theta}
    E_T = H & = & \frac{1}{2}\left[ p_\rho^2 +
\frac{p_\theta^2}{(\rho + 1-f)^2} \right]      \nonumber  \\*
            & + & \frac{\rho^2}{2} - (\rho + 1-f)f \cos\theta .
\end{eqnarray} 

In \cite{C.deSousa2018EnergyDistrib}, we split the spring pendulum Hamiltonian into the spring, pendulum, and coupling energy components. The spring energy $E_S$ characterizes a spring mass moving vertically under the action of gravity:
\begin{equation} \label{eq:ES}
    E_S = \frac{p_\rho^2 + \rho^2}{2} - (\rho + 1-f)f .
\end{equation}
The pendulum energy $E_P$ describes the motion of a simple pendulum and is given by:
\begin{equation} \label{eq:EP}
    E_P = \frac{p_\theta^2}{2} - f \cos \theta .
\end{equation}
The coupling energy $E_C$ arises from the nonlinear coupling between the spring and pendulum oscillations. It is the only energy component that depends on both $\rho$ and $\theta$ associated with the spring and pendulum respectively:
\begin{eqnarray} \label{eq:EC}
    E_C & = &  \frac{p_\theta^2}{2} \left[\frac{ 1}{ (\rho + 1-f)^2} -1\right]       \nonumber  \\*
        & - & (\rho-f)f \cos\theta + (\rho + 1-f)f .
\end{eqnarray}
From expressions (\ref{eq:H_rho_theta})-(\ref{eq:EC}), the total energy of the spring pendulum is given by
\begin{equation}
    E_T = E_S + E_P + E_C.
\end{equation}

These energy components are compatible with the spring and pendulum oscillations and account for their coupling. In Ref. \cite{C.deSousa2018EnergyDistrib}, we applied the analytical expressions (\ref{eq:ES})-(\ref{eq:EC}) to determine how the total energy is shared among the three components for individual trajectories. We verified that the energy components accurately describe the behavior of all kinds of trajectory: periodic, quasi-periodic and chaotic. By calculating phase space and time averages for a great number of trajectories, we also obtained the average energy distribution as a function of total energy and control parameter $f$. Once again, the average energy distribution is a good representation of the global dynamics in phase space for all possible values of $E_T$ and $f$.

In this paper, we investigate how the energy is transferred among the three energy components as the system evolves in time. To do so, we consider the rate of energy exchange
\begin{equation} \label{eq:Pi}
    P_i = \left| \frac{dE_i}{dt} \right| ,
\end{equation}
where $P_i$ corresponds to the power associated with each energy component (\ref{eq:ES})-(\ref{eq:EC}), and the index $i$ represents the type of oscillation: spring $S$, pendulum $P$, and coupled $C$.

\begin{figure}[!t]
    \centering
    \includegraphics[width=1.0\linewidth]{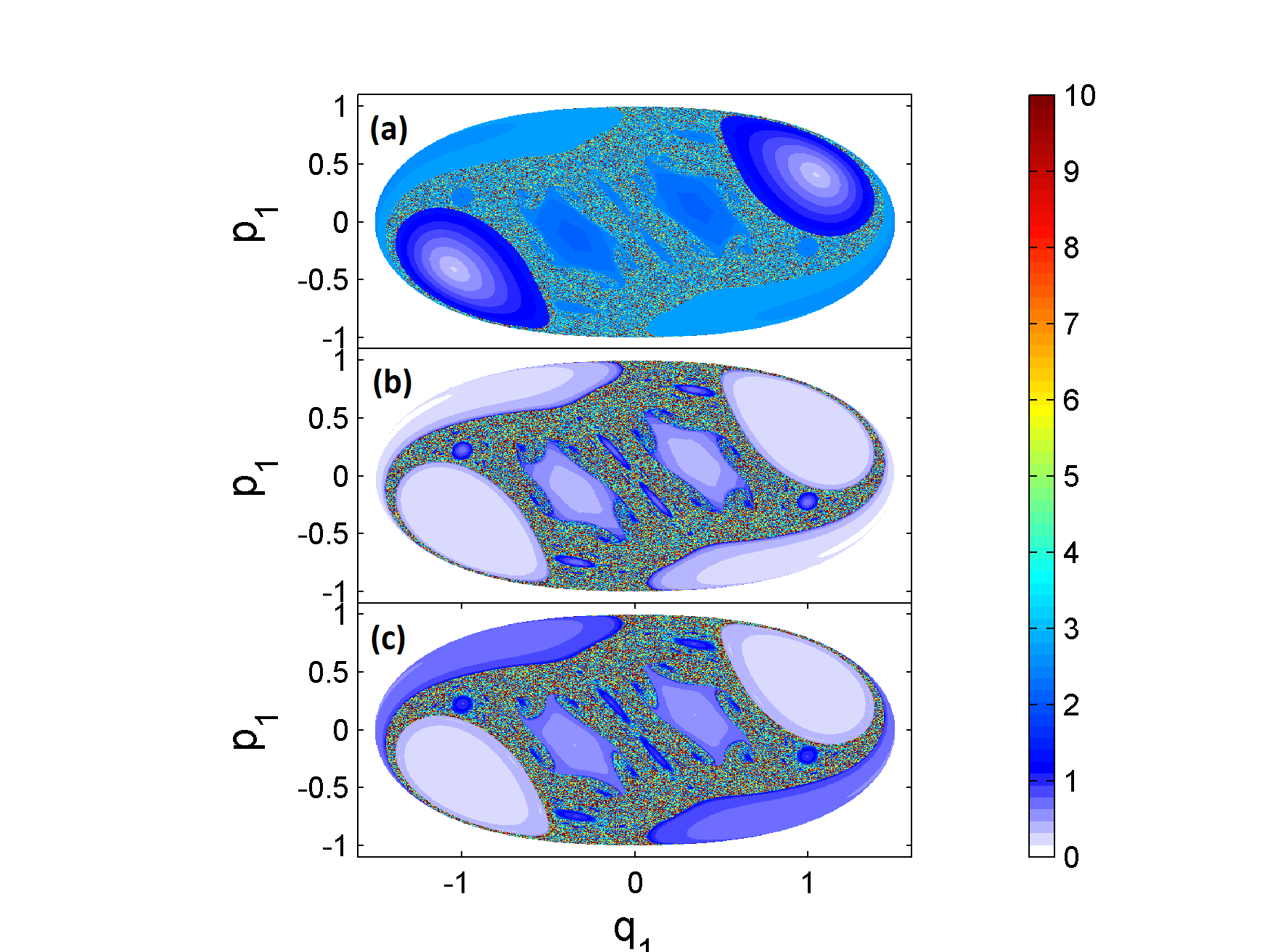}
    \caption{(Color online) Color scale showing the maximum power distribution in the section $p_1 \times q_1$ for $E_T=0.3$ and $f=0.25$: (a) spring power $P_{S,max}$, (b) pendular power $P_{P,max}$, and (c) coupling power $P_{C,max}$.}
    \label{fig:potm03025}
\end{figure}

Through the power terms (\ref{eq:Pi}), we investigate how the total energy is exchanged among the different types of oscillation that an orbit goes through during its time evolution. In Figure \ref{fig:potm03025}, we consider the two-dimensional section $p_1$ {\it vs.} $q_1$ (with $q_2 = 0$) of phase space, and plot in color scale the maximum power $P_{i,max}$ obtained for a long period of time, $t_f=10^3$, in a grid of $600 \times 600$ initial conditions. By comparing Figure \ref{fig:potm03025} with the Poincar\'e section of Figure \ref{fig:ps03025}, we notice that regular and chaotic regions are clearly distinguished from each other by their different values of maximum power. Regular trajectories have well defined values of maximum power, whereas initial conditions in the chaotic sea present non-uniform values of $P_{i,max}$.

The maximum power values of chaotic orbits are typically higher than those of regular orbits for all types of oscillation: spring, pendulum and coupled. It means that energy is exchanged at high rates among the three types of oscillation when the orbit behaves chaotically. On the other hand, energy exchanges in regular trajectories occur at a much lower rate. Furthermore, we observe that the maximum pendulum power is almost null in the six main islands of regular behavior in Figure \ref{fig:potm03025}. Thus, in these regions, energy is exchanged mostly between the spring and coupling energy components.

\begin{figure}[!t]
    \centering
    \includegraphics[width=\linewidth]{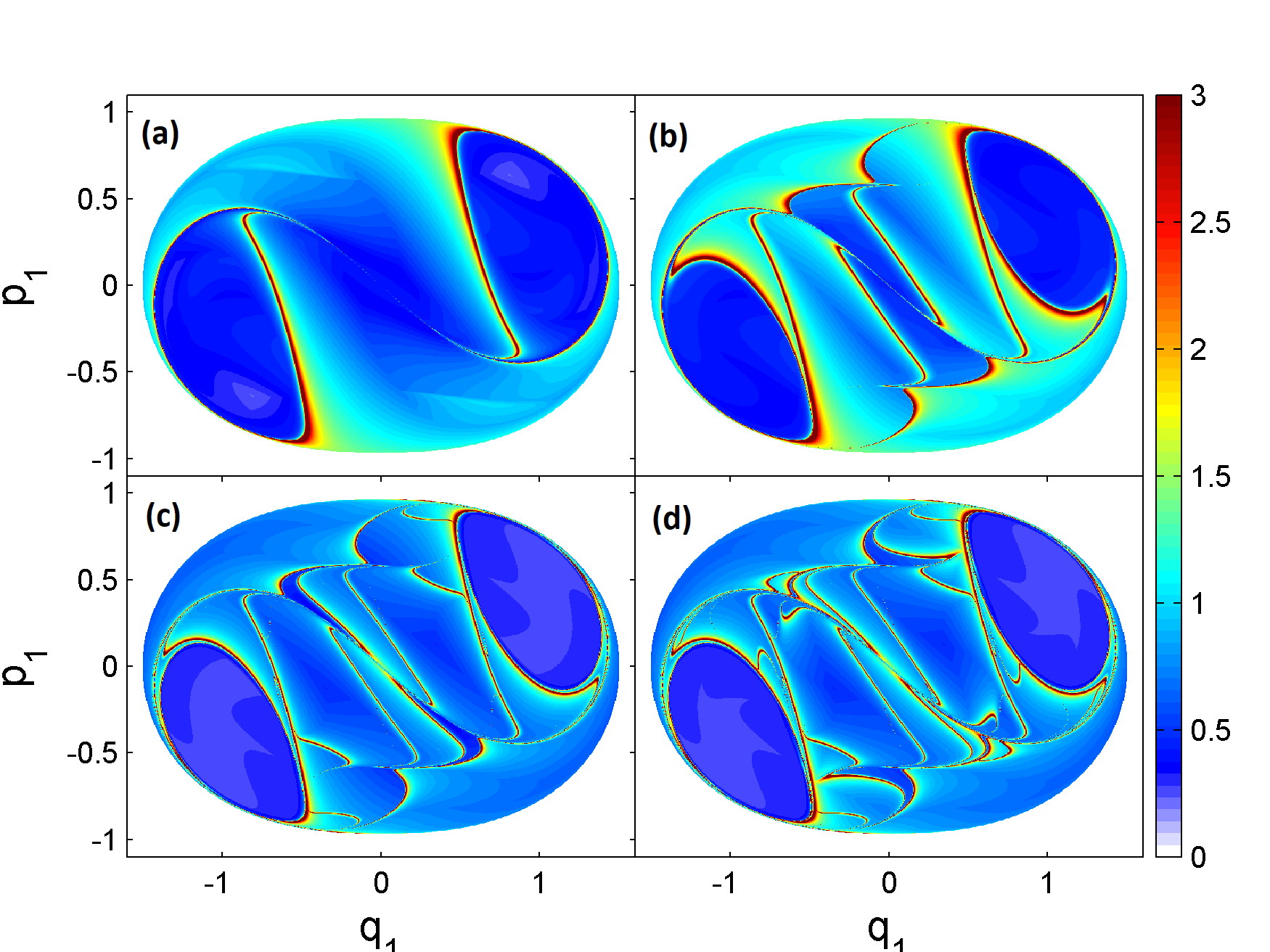}
    \caption{(Color online) Time evolution of the maximum coupling power for $t_i=0$ and (a) $t_f=10$, (b) $t_f=20$, (c) $t_f=30$ and (d) $t_f=40$.} 
    \label{fig:potev03025}
\end{figure}

\begin{figure}[!t]
    \centering
    \includegraphics[width=\linewidth]{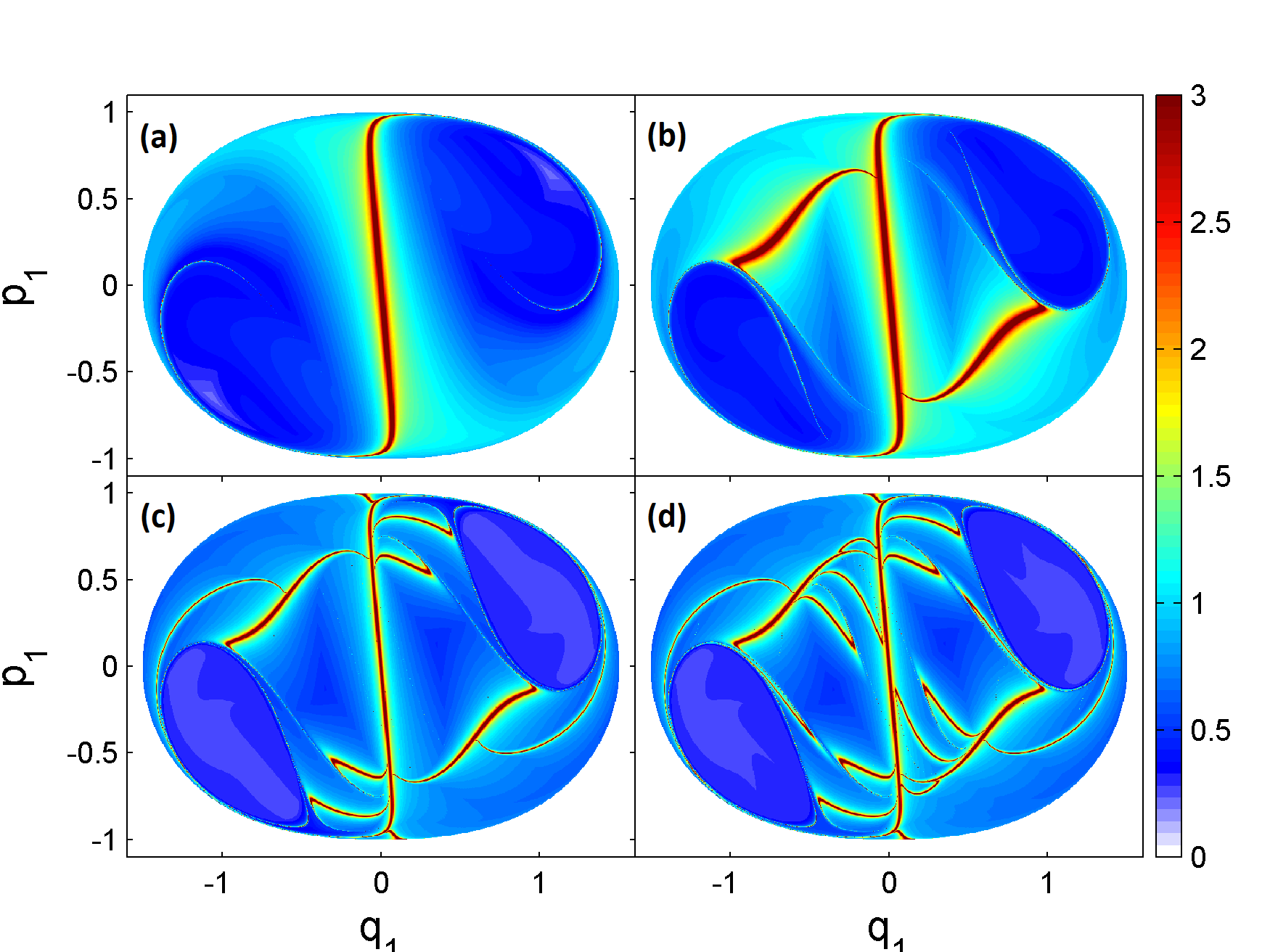}
    \caption{(Color online) Backward time evolution of the maximum coupling power for $t_i=0$ and (a) $t_f=-10$, (b) $t_f=-20$, (c) $t_f=-30$ and (d) $t_f=-40$.}
    \label{fig:potbackev03025}
\end{figure}

From now on, we will focus on the coupling power. To better understand the evolution of $P_{C,max}$ over time, we show in Figure \ref{fig:potev03025} the maximum value of coupling power integrated over different values of final time $t_f$. As time passes from panels (a) to (d), we observe the growth of the red filaments that correspond to the highest values of maximum coupling power ($2.5 \le P_{C,max} \le 3.0$).

\begin{figure}[!t]
    \centering
    \includegraphics[width=1.0\linewidth]{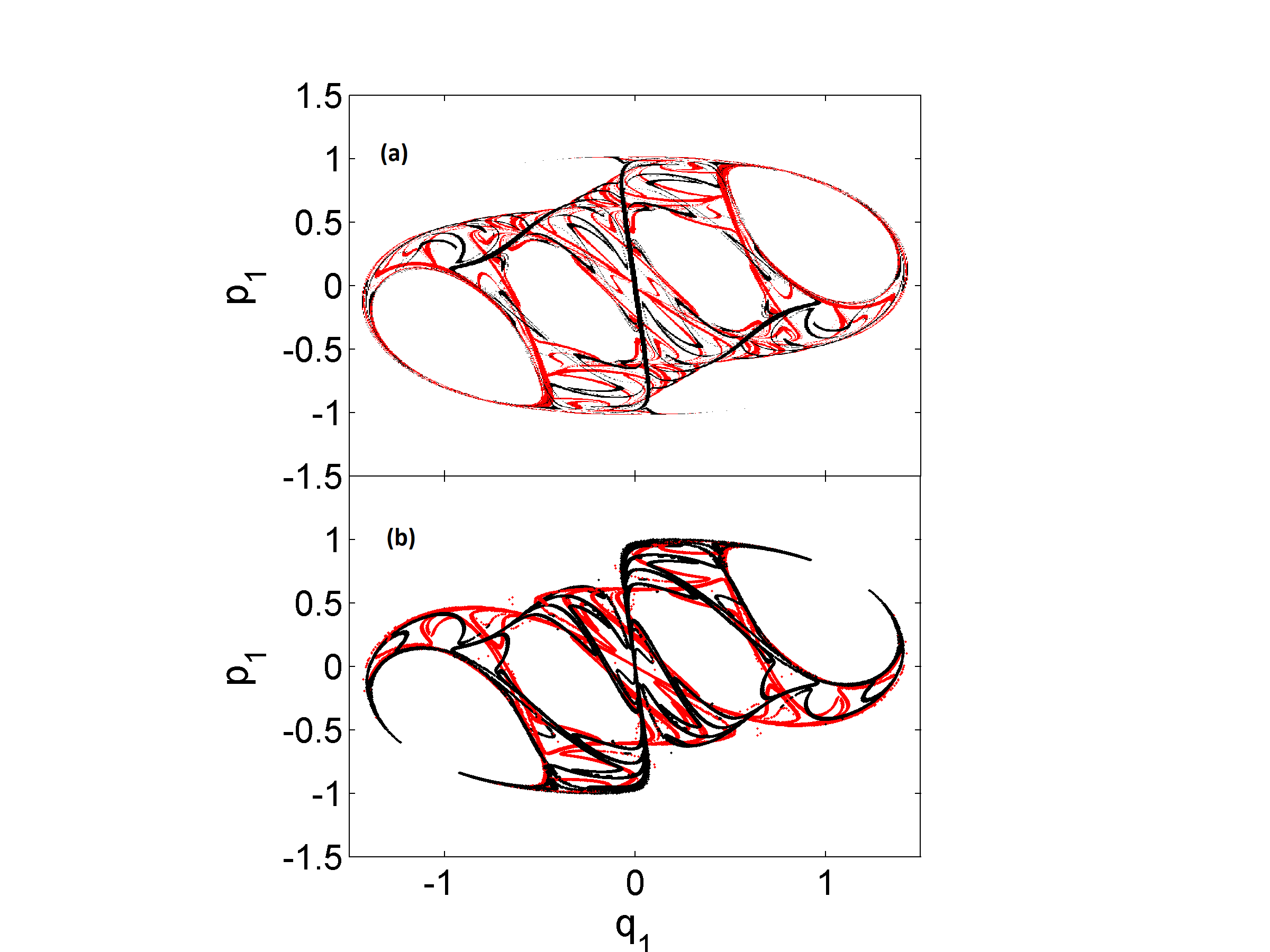}
    \caption{(Color online) (a) Filaments with highest values of maximum coupling power for $t_f = \pm 70$ obtained by integrating the equations of motion forward (red) and backwards (black). (b) Unstable (red) and stable (black) manifolds of the primary hyperbolic point at $p_1 = q_1 = 0$.}
    \label{fig:potfwbackev03025}
\end{figure}

By integrating the equations of motion backwards in time, we obtain another set of red filaments representing the highest values of maximum coupling power, as shown in Figure \ref{fig:potbackev03025}. We combine both Figures \ref{fig:potev03025} and \ref{fig:potbackev03025} to exhibit in Figure \ref{fig:potfwbackev03025}(a) the filaments with the highest values of maximum coupling power for $t_f = \pm 70$. In this figure, the red filaments were calculated by integrating the equations of motion forward in time, whereas the black filaments show the highest values of $P_{C,max}$ for the backward time evolution.

We also obtain the manifolds of the primary hyperbolic point $q_1 = p_1 = 0$, located in the chaotic region of Figure \ref{fig:ps03025}, and compare them with the filaments of highest values of maximum coupling power. To calculate approximations for the invariant manifolds, we apply one standard numerical procedure \cite{Kantz1985RepellersTransients}: we consider a disk of very small radius ($10^{-6}$) centered at the hyperbolic point $(q_1, p_1) = (0,0)$, and containing a large number of initial conditions. Each initial condition belonging to the disk is numerically integrated from $t_i=0$ to $t_f=70$, generating a continuous flow. In Figure \ref{fig:potfwbackev03025}(b), we plot in the section $p_1 \times q_1$ the intersections of this continuous flow with the plane $q_2=0$ whenever $p_2 > 0$. The plot (in red) turns out to be a spaghetti-like region following the unstable manifold that emanates from the periodic orbit with constant $(q_1, p_1) = (0,0)$.

We also integrate the equations of motion backwards in time for the initial conditions within the tiny disk for $t_i=0$ and $t_f = -70$. The plot in section $p_1 \times q_1$ [shown in black in Figure \ref{fig:potfwbackev03025}(b)] is another spaghetti-like region now following the stable manifold of the primary hyperbolic point at $(q_1, p_1) = (0,0)$. In Figure \ref{fig:potfwbackev03025}(b), we display the superposition of the primary unstable (in red) and stable (in black) manifolds for the Poincar\'e section of Figure \ref{fig:ps03025}.

Comparing Figures \ref{fig:potfwbackev03025}(a) and \ref{fig:potfwbackev03025}(b), we identify the filaments of highest coupling power with the manifolds of the primary hyperbolic point $(q_1, p_1) = (0,0)$. The filaments obtained for the forward time evolution correspond to the unstable manifold (both in red in Figure \ref{fig:potfwbackev03025}), whereas the filaments for the backward time evolution represent the stable manifold (both in black).

The homoclinic tangle, which is formed by the intersections of the unstable and stable manifolds, is the source of the highest values of coupling power, i.e. it is responsible for the maximum energy exchange rate among the three different types of energy: spring, pendulum and coupling. Indeed, the homoclinic tangle influences all orbits in the chaotic area, which pass several times in its vicinity. In the same way, the homoclinic tangle leads to the higher energy exchange rates observed in the chaotic sea than in the regular regions of Figure \ref{fig:potm03025}. It means that chaotic trajectories exchange energy at much higher rates between spring, pendulum and coupled oscillations, resulting in high values of all power components $P_{i,max}$ as observed in Figure \ref{fig:potm03025}.

In conclusion, we considered the spring pendulum, a paradigm of intrinsically coupled nonlinear systems, to investigate internal energy exchanges among the different kinds of oscillation it may present. We split the total energy among three energy components that accurately describe the spring, pendulum and coupled oscillations \cite{C.deSousa2018EnergyDistrib}. We then obtained analytical expressions for the energy exchange rate, i.e. the power, associated with each energy component.

We applied these expressions to determine the maximum values of the power components for a large number of trajectories. We observed that regular regions in phase space display well defined values of maximum power for all the components. On the other hand, chaotic areas can be distinguished by their non-uniform values of maximum power. Furthermore, chaotic trajectories generally present higher values of power than regular trajectories.

To determine the reason for the observed features, we analyzed the time evolution of the maximum coupling power in phase space, and identified the trajectories with the highest values of coupling power. We also computed the unstable and stable manifolds of the primary hyperbolic point of the system, and found that they correspond to the trajectories presenting the highest values of coupling power. It means that the coupling power is maximum along the primary manifolds and especially along the homoclinic tangle formed by the intersections of the unstable and stable manifolds. Since chaotic trajectories pass frequently in the vicinity of the homoclinic tangle, the rate of internal energy exchange is much higher for these trajectories.

We confirmed for several values of total energy and control parameter that the maximum coupling power occurs along the manifolds of the primary hyperbolic point, and that all the power components are typically lower for regular trajectories than for chaotic ones. In a future work, we will investigate internal energy exchanges for other nonlinear coupled Hamiltonian systems in order to generalize the results presented in this paper, which may be a useful and efficient way to characterize chaotic orbits and to compute manifolds.

We thank Y. Elskens and A. M. Ozorio de Almeida for fruitful discussions and careful reading of the manuscript. This project has received funding from S\~ao Paulo Research Foundation (FAPESP) under Grant Nos. 2015/05186-0, 2018/03211-6, and 2022/04251-7; Conselho Nacional de Desenvolvimento Cient\'ifico e Tecnol\'ogico (CNPq) under Grant Nos. 407299/2018-1, 302665/2017-0, 403120/2021-7, and 301019/2019-3, and the European Union’s Horizon 2020 research and innovation programme under the Marie Skłodowska-Curie Grant agreement No. 899987.

\bibliographystyle{aipnum4-2}
\bibliography{references.bib}

\begin{thebibliography}{32}%
\makeatletter
\providecommand \@ifxundefined [1]{%
 \@ifx{#1\undefined}
}%
\providecommand \@ifnum [1]{%
 \ifnum #1\expandafter \@firstoftwo
 \else \expandafter \@secondoftwo
 \fi
}%
\providecommand \@ifx [1]{%
 \ifx #1\expandafter \@firstoftwo
 \else \expandafter \@secondoftwo
 \fi
}%
\providecommand \natexlab [1]{#1}%
\providecommand \enquote  [1]{``#1''}%
\providecommand \bibnamefont  [1]{#1}%
\providecommand \bibfnamefont [1]{#1}%
\providecommand \citenamefont [1]{#1}%
\providecommand \href@noop [0]{\@secondoftwo}%
\providecommand \href [0]{\begingroup \@sanitize@url \@href}%
\providecommand \@href[1]{\@@startlink{#1}\@@href}%
\providecommand \@@href[1]{\endgroup#1\@@endlink}%
\providecommand \@sanitize@url [0]{\catcode `\\12\catcode `\$12\catcode
  `\&12\catcode `\#12\catcode `\^12\catcode `\_12\catcode `\%12\relax}%
\providecommand \@@startlink[1]{}%
\providecommand \@@endlink[0]{}%
\providecommand \url  [0]{\begingroup\@sanitize@url \@url }%
\providecommand \@url [1]{\endgroup\@href {#1}{\urlprefix }}%
\providecommand \urlprefix  [0]{URL }%
\providecommand \Eprint [0]{\href }%
\providecommand \doibase [0]{https://doi.org/}%
\providecommand \selectlanguage [0]{\@gobble}%
\providecommand \bibinfo  [0]{\@secondoftwo}%
\providecommand \bibfield  [0]{\@secondoftwo}%
\providecommand \translation [1]{[#1]}%
\providecommand \BibitemOpen [0]{}%
\providecommand \bibitemStop [0]{}%
\providecommand \bibitemNoStop [0]{.\EOS\space}%
\providecommand \EOS [0]{\spacefactor3000\relax}%
\providecommand \BibitemShut  [1]{\csname bibitem#1\endcsname}%
\let\auto@bib@innerbib\@empty
\bibitem [{\citenamefont {Ford}(1992)}]{Ford1992FermiPastaUlam}%
  \BibitemOpen
  \bibfield  {author} {\bibinfo {author} {\bibfnamefont {J.}~\bibnamefont
  {Ford}},\ }\href@noop {} {\bibfield  {journal} {\bibinfo  {journal} {Physics
  Reports}\ }\textbf {\bibinfo {volume} {213}},\ \bibinfo {pages} {271}
  (\bibinfo {year} {1992})}\BibitemShut {NoStop}%
\bibitem [{\citenamefont {Dauxois}, \citenamefont {Peyrard},\ and\
  \citenamefont {Ruffo}(2005)}]{Dauxois2005FermiPastaUlam}%
  \BibitemOpen
  \bibfield  {author} {\bibinfo {author} {\bibfnamefont {T.}~\bibnamefont
  {Dauxois}}, \bibinfo {author} {\bibfnamefont {M.}~\bibnamefont {Peyrard}},\
  and\ \bibinfo {author} {\bibfnamefont {S.}~\bibnamefont {Ruffo}},\
  }\href@noop {} {\bibfield  {journal} {\bibinfo  {journal} {European Journal
  of Physics}\ }\textbf {\bibinfo {volume} {26}},\ \bibinfo {pages} {S3}
  (\bibinfo {year} {2005})}\BibitemShut {NoStop}%
\bibitem [{\citenamefont {Dauxois}(2008)}]{tsingou}%
  \BibitemOpen
  \bibfield  {author} {\bibinfo {author} {\bibfnamefont {T.}~\bibnamefont
  {Dauxois}},\ }\href@noop {} {\bibfield  {journal} {\bibinfo  {journal}
  {Physics Today}\ }\textbf {\bibinfo {volume} {61}},\ \bibinfo {pages} {55}
  (\bibinfo {year} {2008})}\BibitemShut {NoStop}%
\bibitem [{\citenamefont {Vitt}\ and\ \citenamefont
  {Gorelik}(1933)}]{Vitt1933OscillationsSystems}%
  \BibitemOpen
  \bibfield  {author} {\bibinfo {author} {\bibfnamefont {A.}~\bibnamefont
  {Vitt}}\ and\ \bibinfo {author} {\bibfnamefont {G.}~\bibnamefont {Gorelik}},\
  }\href@noop {} {\bibfield  {journal} {\bibinfo  {journal} {Journal of
  Technical Physics}\ }\textbf {\bibinfo {volume} {3}},\ \bibinfo {pages} {294}
  (\bibinfo {year} {1933})}\BibitemShut {NoStop}%
\bibitem [{\citenamefont {N{\'{u}}{\~{n}}ez-Y{\'{e}}pez}\ \emph
  {et~al.}(1990)\citenamefont {N{\'{u}}{\~{n}}ez-Y{\'{e}}pez}, \citenamefont
  {Salas-Brito}, \citenamefont {Vargas},\ and\ \citenamefont
  {Vicente}}]{Nunez-Yepez1990OnsetPendulum}%
  \BibitemOpen
  \bibfield  {author} {\bibinfo {author} {\bibfnamefont {H.~N.}\ \bibnamefont
  {N{\'{u}}{\~{n}}ez-Y{\'{e}}pez}}, \bibinfo {author} {\bibfnamefont {A.~L.}\
  \bibnamefont {Salas-Brito}}, \bibinfo {author} {\bibfnamefont {C.~A.}\
  \bibnamefont {Vargas}},\ and\ \bibinfo {author} {\bibfnamefont
  {L.}~\bibnamefont {Vicente}},\ }\href@noop {} {\bibfield  {journal} {\bibinfo
   {journal} {Physics Letters A}\ }\textbf {\bibinfo {volume} {145}},\ \bibinfo
  {pages} {101} (\bibinfo {year} {1990})}\BibitemShut {NoStop}%
\bibitem [{\citenamefont {Cuerno}, \citenamefont {Ra{\~{n}}ada},\ and\
  \citenamefont {Ruiz-Lorenzo}(1992)}]{Cuerno_AmJP92}%
  \BibitemOpen
  \bibfield  {author} {\bibinfo {author} {\bibfnamefont {R.}~\bibnamefont
  {Cuerno}}, \bibinfo {author} {\bibfnamefont {A.~F.}\ \bibnamefont
  {Ra{\~{n}}ada}},\ and\ \bibinfo {author} {\bibfnamefont {J.~J.}\ \bibnamefont
  {Ruiz-Lorenzo}},\ }\href@noop {} {\bibfield  {journal} {\bibinfo  {journal}
  {American Journal of Physics}\ }\textbf {\bibinfo {volume} {60}},\ \bibinfo
  {pages} {73} (\bibinfo {year} {1992})}\BibitemShut {NoStop}%
\bibitem [{\citenamefont {Carretero-Gonz{\'{a}}lez}, \citenamefont
  {N{\'{u}}{\~{n}}ez-Y{\'{e}}pez},\ and\ \citenamefont
  {Salas-Brito}(1994)}]{Gonzalez_EJP94}%
  \BibitemOpen
  \bibfield  {author} {\bibinfo {author} {\bibfnamefont {R.}~\bibnamefont
  {Carretero-Gonz{\'{a}}lez}}, \bibinfo {author} {\bibfnamefont {H.~N.}\
  \bibnamefont {N{\'{u}}{\~{n}}ez-Y{\'{e}}pez}},\ and\ \bibinfo {author}
  {\bibfnamefont {A.~L.}\ \bibnamefont {Salas-Brito}},\ }\href@noop {}
  {\bibfield  {journal} {\bibinfo  {journal} {European Journal of Physics}\
  }\textbf {\bibinfo {volume} {15}},\ \bibinfo {pages} {139} (\bibinfo {year}
  {1994})}\BibitemShut {NoStop}%
\bibitem [{\citenamefont {van~der Weele}\ and\ \citenamefont
  {de~Kleine}(1996)}]{Weele_PhysA1996}%
  \BibitemOpen
  \bibfield  {author} {\bibinfo {author} {\bibfnamefont {J.~P.}\ \bibnamefont
  {van~der Weele}}\ and\ \bibinfo {author} {\bibfnamefont {E.}~\bibnamefont
  {de~Kleine}},\ }\href@noop {} {\bibfield  {journal} {\bibinfo  {journal}
  {Physica A}\ }\textbf {\bibinfo {volume} {228}},\ \bibinfo {pages} {245}
  (\bibinfo {year} {1996})}\BibitemShut {NoStop}%
\bibitem [{\citenamefont {Kane}\ and\ \citenamefont
  {Kahn}(1968)}]{Kane_JAM1968}%
  \BibitemOpen
  \bibfield  {author} {\bibinfo {author} {\bibfnamefont {T.~R.}\ \bibnamefont
  {Kane}}\ and\ \bibinfo {author} {\bibfnamefont {M.~E.}\ \bibnamefont
  {Kahn}},\ }\href@noop {} {\bibfield  {journal} {\bibinfo  {journal} {Journal
  of Applied Mechanics}\ }\textbf {\bibinfo {volume} {35}},\ \bibinfo {pages}
  {547} (\bibinfo {year} {1968})}\BibitemShut {NoStop}%
\bibitem [{\citenamefont {Tsel'man}(1970)}]{Tselman_JAMM1970}%
  \BibitemOpen
  \bibfield  {author} {\bibinfo {author} {\bibfnamefont {F.~K.}\ \bibnamefont
  {Tsel'man}},\ }\href@noop {} {\bibfield  {journal} {\bibinfo  {journal}
  {Journal of Applied Mathematics and Mechanics}\ }\textbf {\bibinfo {volume}
  {34}},\ \bibinfo {pages} {916} (\bibinfo {year} {1970})}\BibitemShut
  {NoStop}%
\bibitem [{\citenamefont {Rusbridge}(1980)}]{Rusbridge_AmJP1980}%
  \BibitemOpen
  \bibfield  {author} {\bibinfo {author} {\bibfnamefont {M.~G.}\ \bibnamefont
  {Rusbridge}},\ }\href@noop {} {\bibfield  {journal} {\bibinfo  {journal}
  {American Journal of Physics}\ }\textbf {\bibinfo {volume} {48}},\ \bibinfo
  {pages} {146} (\bibinfo {year} {1980})}\BibitemShut {NoStop}%
\bibitem [{\citenamefont {Breitenberger}\ and\ \citenamefont
  {Mueller}(1981)}]{Breitenberger_JMP1981}%
  \BibitemOpen
  \bibfield  {author} {\bibinfo {author} {\bibfnamefont {E.}~\bibnamefont
  {Breitenberger}}\ and\ \bibinfo {author} {\bibfnamefont {R.~D.}\ \bibnamefont
  {Mueller}},\ }\href@noop {} {\bibfield  {journal} {\bibinfo  {journal}
  {Journal of Mathematical Physics}\ }\textbf {\bibinfo {volume} {22}},\
  \bibinfo {pages} {1196} (\bibinfo {year} {1981})}\BibitemShut {NoStop}%
\bibitem [{\citenamefont {Lai}(1984)}]{Lai_AmJP1984}%
  \BibitemOpen
  \bibfield  {author} {\bibinfo {author} {\bibfnamefont {H.~M.}\ \bibnamefont
  {Lai}},\ }\href@noop {} {\bibfield  {journal} {\bibinfo  {journal} {American
  Journal of Physics}\ }\textbf {\bibinfo {volume} {52}},\ \bibinfo {pages}
  {219} (\bibinfo {year} {1984})}\BibitemShut {NoStop}%
\bibitem [{\citenamefont {Fermi}\ and\ \citenamefont
  {Rasetti}(1931)}]{Fermi1931UberSteinsalzes}%
  \BibitemOpen
  \bibfield  {author} {\bibinfo {author} {\bibfnamefont {E.}~\bibnamefont
  {Fermi}}\ and\ \bibinfo {author} {\bibfnamefont {F.}~\bibnamefont
  {Rasetti}},\ }\href@noop {} {\bibfield  {journal} {\bibinfo  {journal}
  {Zeitschrift f{\"{u}}r Physik}\ }\textbf {\bibinfo {volume} {71}},\ \bibinfo
  {pages} {689} (\bibinfo {year} {1931})}\BibitemShut {NoStop}%
\bibitem [{\citenamefont {Amat}\ and\ \citenamefont
  {Pimbert}(1965)}]{Amat1965OnDioxide}%
  \BibitemOpen
  \bibfield  {author} {\bibinfo {author} {\bibfnamefont {G.}~\bibnamefont
  {Amat}}\ and\ \bibinfo {author} {\bibfnamefont {M.}~\bibnamefont {Pimbert}},\
  }\href@noop {} {\bibfield  {journal} {\bibinfo  {journal} {Journal of
  Molecular Spectroscopy}\ }\textbf {\bibinfo {volume} {16}},\ \bibinfo {pages}
  {278} (\bibinfo {year} {1965})}\BibitemShut {NoStop}%
\bibitem [{\citenamefont {Jacob}, \citenamefont {Gross},\ and\ \citenamefont
  {Dreizler}(1978)}]{Jacob_JPB1978}%
  \BibitemOpen
  \bibfield  {author} {\bibinfo {author} {\bibfnamefont {B.}~\bibnamefont
  {Jacob}}, \bibinfo {author} {\bibfnamefont {E.~K.~U.}\ \bibnamefont
  {Gross}},\ and\ \bibinfo {author} {\bibfnamefont {R.~M.}\ \bibnamefont
  {Dreizler}},\ }\href@noop {} {\bibfield  {journal} {\bibinfo  {journal}
  {Journal of Physics B}\ }\textbf {\bibinfo {volume} {11}},\ \bibinfo {pages}
  {3795} (\bibinfo {year} {1978})}\BibitemShut {NoStop}%
\bibitem [{\citenamefont {Contopoulos}(1963)}]{Contopoulos1963ResonanceI}%
  \BibitemOpen
  \bibfield  {author} {\bibinfo {author} {\bibfnamefont {G.}~\bibnamefont
  {Contopoulos}},\ }\href@noop {} {\bibfield  {journal} {\bibinfo  {journal}
  {The Astronomical Journal}\ }\textbf {\bibinfo {volume} {68}},\ \bibinfo
  {pages} {763} (\bibinfo {year} {1963})}\BibitemShut {NoStop}%
\bibitem [{\citenamefont {Hori}(1966)}]{Hori_ASJ1966}%
  \BibitemOpen
  \bibfield  {author} {\bibinfo {author} {\bibfnamefont {G.-I.}\ \bibnamefont
  {Hori}},\ }\href@noop {} {\bibfield  {journal} {\bibinfo  {journal}
  {Publications of the Astronomical Society of Japan}\ }\textbf {\bibinfo
  {volume} {18}},\ \bibinfo {pages} {287} (\bibinfo {year} {1966})}\BibitemShut
  {NoStop}%
\bibitem [{\citenamefont {Broucke}\ and\ \citenamefont
  {Baxa}(1973)}]{Broucke1973PeriodicSystem}%
  \BibitemOpen
  \bibfield  {author} {\bibinfo {author} {\bibfnamefont {R.}~\bibnamefont
  {Broucke}}\ and\ \bibinfo {author} {\bibfnamefont {P.~A.}\ \bibnamefont
  {Baxa}},\ }\href@noop {} {\bibfield  {journal} {\bibinfo  {journal}
  {Celestial Mechanics}\ }\textbf {\bibinfo {volume} {8}},\ \bibinfo {pages}
  {261} (\bibinfo {year} {1973})}\BibitemShut {NoStop}%
\bibitem [{\citenamefont {Hitzl}(1975)}]{Hitzl_CM1975}%
  \BibitemOpen
  \bibfield  {author} {\bibinfo {author} {\bibfnamefont {D.~L.}\ \bibnamefont
  {Hitzl}},\ }\href@noop {} {\bibfield  {journal} {\bibinfo  {journal}
  {Celestial Mechanics}\ }\textbf {\bibinfo {volume} {12}},\ \bibinfo {pages}
  {359} (\bibinfo {year} {1975})}\BibitemShut {NoStop}%
\bibitem [{\citenamefont {Sagdeev}\ and\ \citenamefont
  {Galeev}(1969)}]{Sagdeev1969NonlinearTheory}%
  \BibitemOpen
  \bibfield  {author} {\bibinfo {author} {\bibfnamefont {R.~Z.}\ \bibnamefont
  {Sagdeev}}\ and\ \bibinfo {author} {\bibfnamefont {A.~A.}\ \bibnamefont
  {Galeev}},\ }\href@noop {} {\emph {\bibinfo {title} {{Nonlinear plasma
  theory}}}}\ (\bibinfo  {publisher} {Benjamin},\ \bibinfo {address} {New
  York},\ \bibinfo {year} {1969})\ \bibinfo {note} {chap. I}\BibitemShut
  {NoStop}%
\bibitem [{\citenamefont {Horton}(2012)}]{Horton2012TurbulentPlasmas}%
  \BibitemOpen
  \bibfield  {author} {\bibinfo {author} {\bibfnamefont {W.}~\bibnamefont
  {Horton}},\ }\href@noop {} {\emph {\bibinfo {title} {{Turbulent transport in
  Magnetized Plasmas}}}}\ (\bibinfo  {publisher} {World Scientific Publishing
  Co.},\ \bibinfo {address} {Singapore},\ \bibinfo {year} {2012})\BibitemShut
  {NoStop}%
\bibitem [{\citenamefont {Armstrong}\ \emph {et~al.}(1962)\citenamefont
  {Armstrong}, \citenamefont {Bloembergen}, \citenamefont {Ducuing},\ and\
  \citenamefont {Pershan}}]{Armstrong_PR1962}%
  \BibitemOpen
  \bibfield  {author} {\bibinfo {author} {\bibfnamefont {J.~A.}\ \bibnamefont
  {Armstrong}}, \bibinfo {author} {\bibfnamefont {N.}~\bibnamefont
  {Bloembergen}}, \bibinfo {author} {\bibfnamefont {J.}~\bibnamefont
  {Ducuing}},\ and\ \bibinfo {author} {\bibfnamefont {P.~S.}\ \bibnamefont
  {Pershan}},\ }\href@noop {} {\bibfield  {journal} {\bibinfo  {journal}
  {Physical Review}\ }\textbf {\bibinfo {volume} {127}},\ \bibinfo {pages}
  {1918} (\bibinfo {year} {1962})}\BibitemShut {NoStop}%
\bibitem [{\citenamefont {Holmes}\ \emph {et~al.}(2006)\citenamefont {Holmes},
  \citenamefont {Full}, \citenamefont {Koditschek},\ and\ \citenamefont
  {Guckenheimer}}]{Holmes2006TheChallenges}%
  \BibitemOpen
  \bibfield  {author} {\bibinfo {author} {\bibfnamefont {P.}~\bibnamefont
  {Holmes}}, \bibinfo {author} {\bibfnamefont {R.~J.}\ \bibnamefont {Full}},
  \bibinfo {author} {\bibfnamefont {D.}~\bibnamefont {Koditschek}},\ and\
  \bibinfo {author} {\bibfnamefont {J.}~\bibnamefont {Guckenheimer}},\
  }\href@noop {} {\bibfield  {journal} {\bibinfo  {journal} {SIAM Review}\
  }\textbf {\bibinfo {volume} {48}},\ \bibinfo {pages} {207} (\bibinfo {year}
  {2006})}\BibitemShut {NoStop}%
\bibitem [{\citenamefont {Anh}\ \emph {et~al.}(2007)\citenamefont {Anh},
  \citenamefont {Matsuhisa}, \citenamefont {Viet},\ and\ \citenamefont
  {Yasuda}}]{Anh2007VibrationAbsorber}%
  \BibitemOpen
  \bibfield  {author} {\bibinfo {author} {\bibfnamefont {N.~D.}\ \bibnamefont
  {Anh}}, \bibinfo {author} {\bibfnamefont {H.}~\bibnamefont {Matsuhisa}},
  \bibinfo {author} {\bibfnamefont {L.~D.}\ \bibnamefont {Viet}},\ and\
  \bibinfo {author} {\bibfnamefont {M.}~\bibnamefont {Yasuda}},\ }\href@noop {}
  {\bibfield  {journal} {\bibinfo  {journal} {Journal of Sound and Vibration}\
  }\textbf {\bibinfo {volume} {307}},\ \bibinfo {pages} {187} (\bibinfo {year}
  {2007})}\BibitemShut {NoStop}%
\bibitem [{\citenamefont {Wang}, \citenamefont {Li},\ and\ \citenamefont
  {Xie}(2011)}]{Wang2011}%
  \BibitemOpen
  \bibfield  {author} {\bibinfo {author} {\bibfnamefont {D.}~\bibnamefont
  {Wang}}, \bibinfo {author} {\bibfnamefont {J.}~\bibnamefont {Li}},\ and\
  \bibinfo {author} {\bibfnamefont {Q.}~\bibnamefont {Xie}},\ }\href@noop {}
  {\bibfield  {journal} {\bibinfo  {journal} {Advances in Structural
  Engineering}\ }\textbf {\bibinfo {volume} {14}},\ \bibinfo {pages} {445}
  (\bibinfo {year} {2011})}\BibitemShut {NoStop}%
\bibitem [{\citenamefont {Castillo-Rivera}\ and\ \citenamefont
  {Tomas-Rodriguez}(2017)}]{Castillo-Rivera2017}%
  \BibitemOpen
  \bibfield  {author} {\bibinfo {author} {\bibfnamefont {S.}~\bibnamefont
  {Castillo-Rivera}}\ and\ \bibinfo {author} {\bibfnamefont {M.}~\bibnamefont
  {Tomas-Rodriguez}},\ }\href@noop {} {\bibfield  {journal} {\bibinfo
  {journal} {Nonlinear Dynamics}\ }\textbf {\bibinfo {volume} {88}},\ \bibinfo
  {pages} {2933} (\bibinfo {year} {2017})}\BibitemShut {NoStop}%
\bibitem [{\citenamefont {C.~de Sousa}\ \emph {et~al.}(2018)\citenamefont
  {C.~de Sousa}, \citenamefont {Marcus}, \citenamefont {L.~Caldas},\ and\
  \citenamefont {L.~Viana}}]{C.deSousa2018EnergyDistrib}%
  \BibitemOpen
  \bibfield  {author} {\bibinfo {author} {\bibfnamefont {M.}~\bibnamefont
  {C.~de Sousa}}, \bibinfo {author} {\bibfnamefont {F.~A.}\ \bibnamefont
  {Marcus}}, \bibinfo {author} {\bibfnamefont {I.}~\bibnamefont {L.~Caldas}},\
  and\ \bibinfo {author} {\bibfnamefont {R.}~\bibnamefont {L.~Viana}},\
  }\href@noop {} {\bibfield  {journal} {\bibinfo  {journal} {Physica A:
  Statistical Mechanics and its Applications}\ }\textbf {\bibinfo {volume}
  {509}},\ \bibinfo {pages} {1110} (\bibinfo {year} {2018})}\BibitemShut
  {NoStop}%
\bibitem [{\citenamefont {{Ozorio de
  Almeida}}(1988)}]{OzoriodeAlmeida1988HamiltonianSystems}%
  \BibitemOpen
  \bibfield  {author} {\bibinfo {author} {\bibfnamefont {A.~M.}\ \bibnamefont
  {{Ozorio de Almeida}}},\ }\href@noop {} {\emph {\bibinfo {title} {Hamiltonian
  systems: Chaos and quantization}}}\ (\bibinfo  {publisher} {Cambridge
  University Press},\ \bibinfo {address} {Cambridge},\ \bibinfo {year}
  {1988})\BibitemShut {NoStop}%
\bibitem [{\citenamefont {Lichtenberg}\ and\ \citenamefont
  {Lieberman}(1992)}]{Lichtenberg1992}%
  \BibitemOpen
  \bibfield  {author} {\bibinfo {author} {\bibfnamefont {A.~J.}\ \bibnamefont
  {Lichtenberg}}\ and\ \bibinfo {author} {\bibfnamefont {M.~A.}\ \bibnamefont
  {Lieberman}},\ }\href@noop {} {\emph {\bibinfo {title} {Regular and chaotic
  dynamics}}},\ \bibinfo {edition} {2nd}\ ed.\ (\bibinfo  {publisher}
  {Springer},\ \bibinfo {address} {New York},\ \bibinfo {year}
  {1992})\BibitemShut {NoStop}%
\bibitem [{\citenamefont {Meiss}(1992)}]{Meiss1992}%
  \BibitemOpen
  \bibfield  {author} {\bibinfo {author} {\bibfnamefont {J.~D.}\ \bibnamefont
  {Meiss}},\ }\href@noop {} {\bibfield  {journal} {\bibinfo  {journal} {Reviews
  of Modern Physics}\ }\textbf {\bibinfo {volume} {64}},\ \bibinfo {pages}
  {795} (\bibinfo {year} {1992})}\BibitemShut {NoStop}%
\bibitem [{\citenamefont {Kantz}\ and\ \citenamefont
  {Grassberger}(1985)}]{Kantz1985RepellersTransients}%
  \BibitemOpen
  \bibfield  {author} {\bibinfo {author} {\bibfnamefont {H.}~\bibnamefont
  {Kantz}}\ and\ \bibinfo {author} {\bibfnamefont {P.}~\bibnamefont
  {Grassberger}},\ }\href@noop {} {\bibfield  {journal} {\bibinfo  {journal}
  {Physica D: Nonlinear Phenomena}\ }\textbf {\bibinfo {volume} {17}},\
  \bibinfo {pages} {75} (\bibinfo {year} {1985})}\BibitemShut {NoStop}%
\end{thebibliography}%

\end{document}